\documentclass{article}
\usepackage[preprint]{spconf}

\toappear{\begin{minipage}{\textwidth}\footnotesize© 20XX IEEE. Personal use of this material is permitted. Permission from IEEE must be obtained for all other uses, in any current or future media, including reprinting/republishing this material for advertising or promotional purposes, creating new collective works, for resale or redistribution to servers or lists, or reuse of any copyrighted component of this work in other works.\end{minipage}}
\usepackage{spconf,amsmath,graphicx,amsfonts,cite,subcaption,booktabs, multirow}
\usepackage[absolute]{textpos}

\ninept

\usepackage{soul,color}
\usepackage{cite}
\title{SKILL: Similarity-aware Knowledge distILLation for Speech Self-Supervised Learning}

\name{Luca Zampierin$^{1,2}$\sthanks{Work done during an internship at Sony Europe B.V., Stuttgart Lab. 1}, Ghouthi Boukli Hacene$^{1,5}$, Bac Nguyen$^{1}$, Mirco Ravanelli$^{3,4,5}$}
\address{$^1$Sony Europe B.V., Stuttgart Laboratory 1, $^2$Ecole Polytechnique Fédérale de Lausanne, \\ $^3$Concordia University, $^4$Université de Montréal, $^5$Mila-Québec AI Institute\\}

\begin{document}
%
\maketitle
\begin{abstract}

Self-supervised learning (SSL) has achieved remarkable success across various speech-processing tasks. To enhance its efficiency, previous works often leverage the use of compression techniques. 
A notable recent attempt is DPHuBERT, which applies joint knowledge distillation (KD) and structured pruning to learn a significantly smaller SSL model.
In this paper, we contribute to this research domain by introducing SKILL, a novel method that conducts distillation across groups of layers instead of distilling individual arbitrarily selected layers within the teacher network.
The identification of the layers to distill is achieved through a hierarchical clustering procedure applied to layer similarity measures.
Extensive experiments demonstrate that our distilled version of WavLM Base+ not only outperforms DPHuBERT but also achieves state-of-the-art results in the 30M parameters model class across several SUPERB tasks.

\end{abstract}

\section{Introduction}
\label{sec:intro}
Self-supervised speech representations have demonstrated superior performance across multiple tasks~\cite{babu2021xls,baevski2019vq,baevski2020wav2vec,chen2022wavlm,hsu2021hubert,ravanelli2020multi,metricgan,sadhu2021wav2vec,chung2021w2v, zaiem2023benchmarking, Evain_2021,yang2021superb,speechbrain}, including automatic speech recognition (ASR) \cite{hsu2021hubert,chung2021w2v,chen2022wavlm}, automatic speaker verification (ASV) \cite{mutualinfo2019, fan2020exploring}, keyword spotting (KS) \cite{hussain2022multi}, and emotion recognition (ER) \cite{wang2023speech}.  Their applicability in real-world scenarios, however, is still limited by their memory footprint and computational power. To alleviate this drawback, previous works focused on reducing the computational complexity and memory requirements of SSL models ~\cite{lecun1990optimal,hacene2021dnn,courbariaux2015binaryconnect,dettmers2023qlora,zaiem2023fine}. One solution is using knowledge distillation, i.e., training a small network (called \textit{student}) by transferring information from a larger network (called \textit{teacher}). Early works such as DistillHuBERT \cite{chang2022distilhubert} and FitHuBERT \cite{lee2022fithubert} have applied task-agnostic knowledge distillation to HuBERT \cite{hsu2021hubert}. 
Along this line, a recent study demonstrated that varying depth and width of the student model, while keeping the same size, has a significant impact on performance \cite{ashihara2022deep}.
The novel DPHuBERT model leverages these insights. Namely, it learns the optimal student network by jointly performing structured pruning guided by distillation \cite{peng23c_interspeech}.
The method was evaluated on various SSL models, including HuBERT Base  \cite{hsu2021hubert}, WavLM Base+ \cite{chen2022wavlm}, and HuBERT Large \cite{hsu2021hubert}. 
The selection of the layers to be distilled is based on simple heuristics (the authors select layers at a uniform distance, i.e., \{0,4,8,12\} and \{0,8,16,24\} for Base and Large models, respectively).

\begin{figure}
  \centering
  \includegraphics[width=0.43\textwidth]{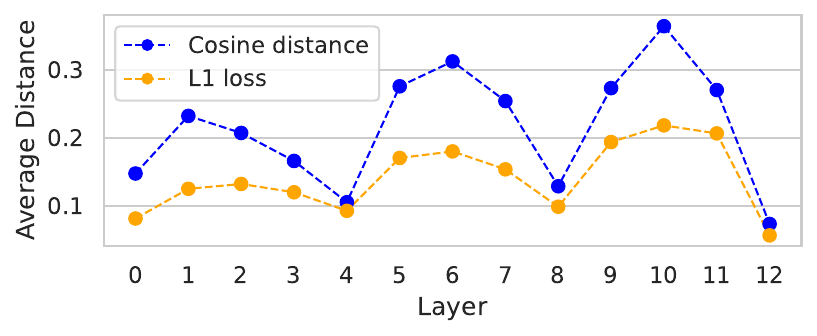}
  \caption{Average cosine and $\ell_1$ distance between WavLM Base+ \cite{chen2022wavlm} and its distilled version DPWavLM \cite{peng23c_interspeech}.}
  \label{fig:DPWavLM_distance}
  \vspace{-0.5cm}
\end{figure}

We argue that the arbitrary selection of layers for distillation is biasing the student network toward the information stored in these layers. For an illustrative purpose, Fig.~\ref{fig:DPWavLM_distance} shows the layer-wise average $\ell_1$ distance (orange) and cosine distance (blue) between WavLM Base+ and its distilled version DPWavLM \cite{peng23c_interspeech} over 200 randomly selected audios from LibriSpeech \cite{panayotov2015librispeech}. 
This result highlights the strong bias of the distilled network towards the selected layers, as evidenced by the consistently larger distances observed across unselected layers compared to the selected ones.
Moreover, identifying the most suitable subset of layers for distillation is computationally expensive and requires careful hyperparameter tuning.
Since each layer can be either distilled or not, the number of combinations increases exponentially with the number of layers. Finally, the layer-to-layer mapping strategy assumes that the same information should be stored at the same level in both teacher and student networks. This assumption might be too restrictive, especially when the size of the student network is significantly reduced~\cite{chen2021distilling}. 

\begin{figure*}[t]
    \begin{subfigure}{0.5\textwidth}
        \centering
        \includegraphics[width=0.55\textwidth]{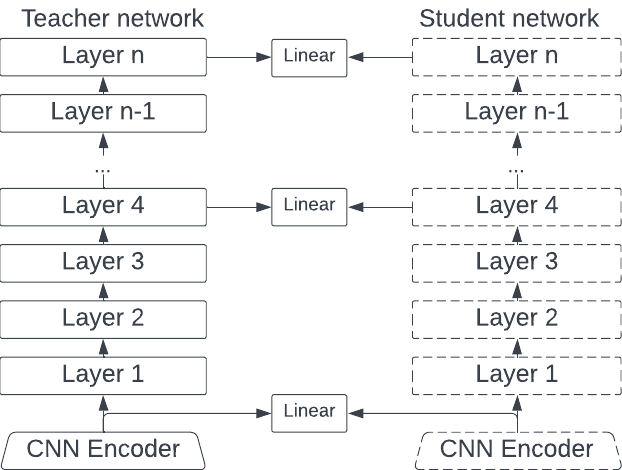}
        \caption{DPHuBERT distillation strategy}
    \end{subfigure}%
    \begin{subfigure}{0.5\textwidth}
        \centering
        \includegraphics[width=0.95\textwidth]{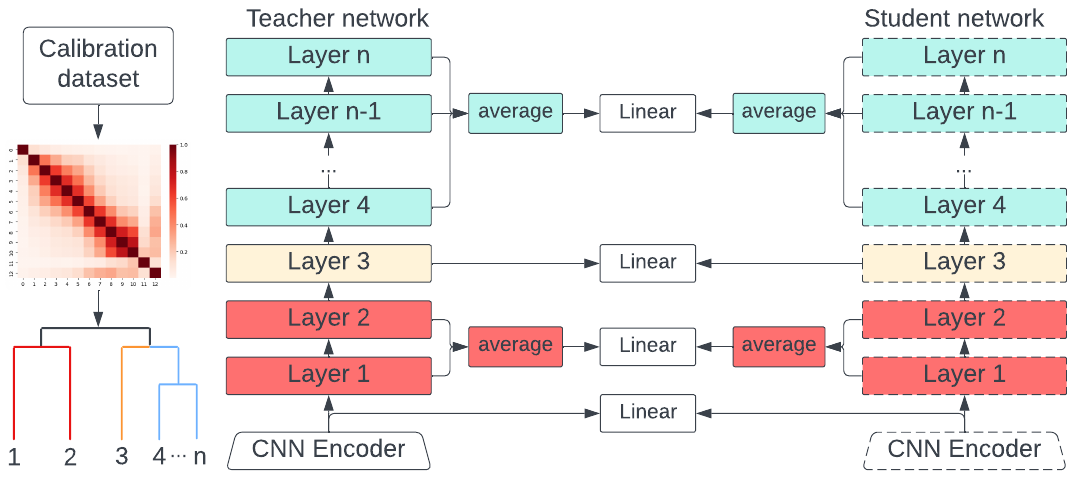}
        \caption{SKILL distillation strategy}
    \end{subfigure}
    \caption{Comparison between DPHuBERT \cite{peng23c_interspeech} and SKILL (ours) distillation strategies. (a) DPHuBERT distills only a set of pre-selected layers. (b) In SKILL, teacher layers are first clustered based on their similarity evaluated on a calibration dataset. The distillation is then performed on average representations of these clusters. (Dashed modules indicate the prunable parameters during stage 1 of training).}
    \label{fig2:summary_method}
    \vspace{-0.5cm}
\end{figure*}

In this paper, we aim to mitigate these limitations with a novel technique called \textbf{S}imilarity-aware \textbf{K}nowledge dist\textbf{ILL}ation (SKILL). Instead of distilling single arbitrarily selected layers, SKILL distills information from all layers. Our approach begins by identifying layers encoding similar information, employing a hierarchical clustering approach based on the similarity between layers in the teacher network. Subsequently, the distillation process occurs on the average representations of these identified layers. Unlike DPHuBERT, our method is data-dependent, which does not require any heuristic selection.

Our experimental validation conducted on the Speech processing Universal PERformance Benchmark (SUPERB)~\cite{yang2021superb} shows that SKILL leads to performance improvements and better generalization capabilities compared to DPHuBERT \cite{peng23c_interspeech}. With SKILL, we can reduce WavLM Base+~\cite{chen2022wavlm} from 94.70M to 23.51M parameters, achieving state-of-the-art results across several SUPERB tasks in the 30M parameters models category. SKILL also demonstrates to be more robust than DPHuBERT to varying sizes of the distilled model. Finally, we compress HuBERT Base \cite{hsu2021hubert} evidencing the model-agnosticism of SKILL.

\section{Background}
\label{sec:preliminaries}

The closest technique to SKILL is DPHuBERT \cite{peng23c_interspeech}, which involves two stages. Initially, the student is initialized from the teacher and jointly distilled and pruned for a smaller network. In the second stage, the student's architecture remains fixed and the parameters are further distilled for better alignment of representations.

\subsection{Knowledge distillation}
In DPHuBERT \cite{peng23c_interspeech}, a multi-stage layer-to-layer feature-based distillation strategy is used, where the layer matching is heuristically pre-defined, as shown in Fig.~\ref{fig2:summary_method}a. The distillation loss is defined as:

\begin{equation}
    \mathcal{L}^{dis} = \sum_{i \in S} \gamma_i \mathcal{L}(\textbf{X}_i^{\text{tea}}, \textbf{X}_i^{\text{stu}} \cdot \textbf{W}_i)
    \label{distillation_loss}
\end{equation}

\noindent where $\textbf{X}_i^{\text{tea}} \in \mathbb{R}^{T \times d_{tea}}$ and $\textbf{X}_i^{\text{stu}} \in \mathbb{R}^{T \times d_{stu}}$ represent the output of the i-th Transformer layer of the teacher and student networks, respectively, $\mathcal{L}$ is the loss function defined as the sum of the $\ell_1$ and cosine distances, $\textbf{W}_i$ is a trainable linear projection, $S$ is the set of indexes indicating the layers to be distilled, and $\gamma_i$ is a layer-wise weighting factor such that $\sum_{i \in S}\gamma_i=1$. Here, $T$ is the sequence length and $d$ represents the embedding dimension. DPHuBERT \cite{peng23c_interspeech} uses $S=\{0,4,8,12\}$ and $S=\{0,8,16,24\}$ for Base and Large models, respectively, with equal weighting factors (i.e. $\forall i \in S, \gamma_i =\frac{1}{4}$). In this paper, we demonstrate that this choice is suboptimal as it leads to a student network with fewer generalization capabilities.

\subsection{Structured pruning}
In the first stage, the student network is progressively pruned in a structured manner until it reaches the pre-defined sparsity target. DPHuBERT \cite{peng23c_interspeech} follows previous studies \cite{wang2020structured,xia2022structured,peng2023structured} in formulating the structured pruning as a $\ell_0$ regularization. 

Let $\theta_j$ represent a set of prunable parameters of the student model (i.e., output channels in CNN layers, attention heads in MHAs layers, or hidden units in FFN) and $n$ the total number of prunable parameter groups. The pruning is modeled using a mask binary variable $z_j \in \{0,1\}$ for each $\theta_j$.
The mask random variable, $\textbf{z}=\{z_j\}_{j=1}^n$, follows a probability distribution $q(\textbf{z}; \boldsymbol{\alpha})$, where $\boldsymbol{\alpha}$ are learnable parameters.
Due to the discrete nature of the masks, it is intractable to update the $\boldsymbol{\alpha}$ parameters via backpropagation. To overcome this issue, the reparameterization trick introduced in \cite{louizos2018learning} is employed, which uses a differentiable Hard Concrete distribution to sample $\textbf{z}$. Moreover, following previous works in NLP \cite{peng2023structured,xia2022structured}, DPHuBERT \cite{peng23c_interspeech} proposes to explicitly control the target sparsity $t$, thus leading to the following constrained optimization problem:

\begin{equation}
\label{eq:pruning}
\begin{aligned}
\min_{\theta, \alpha} \, & \mathbb{E}_{z \sim q}\left[\frac{1}{D} \sum_{k=1}^{D} \mathcal{L}^{\text{dis}}\left(f^{\text{tea}}(\textbf{x}_k), f^{\text{stu}}(\textbf{x}_k; \boldsymbol{\tilde{\theta}})\right)\right], \ \text{s.t.} & s(\boldsymbol{\alpha}) = t
\end{aligned}
\end{equation}

\noindent where $f^{\text{tea}}(.)$ and $f^{\text{stu}}(.)$ represent the frozen teacher and student networks, respectively, $\boldsymbol{\tilde{\theta}}=\{\tilde{\theta_j}\}_{j=1}^n$ represents the pruned parameters of the student model, that is, $\tilde{\theta}_j = \theta_j  z_j$, $D$ is the total number of unlabeled training data, and $s(\boldsymbol{\alpha})$ indicates the current sparsity of the student network, which depends on the $\boldsymbol{\alpha}$ parameters of the Hard Concrete distribution.
In SKILL we inherit the structured pruning approach used in DPHuBERT.

\section{Proposed Method}
\label{sec:Method}
The proposed SKILL method extends DPHuBERT \cite{peng23c_interspeech}, maintaining its two-stage training setup and structured pruning (see Section~\ref{sec:preliminaries}). However, we propose a new KD technique to address the issue of manual layer selection. Fig.~\ref{fig2:summary_method}b illustrates the proposed distillation strategy. Unlike DPHuBERT, SKILL identifies layers with similar information by comparing layer-wise activations with a small calibration dataset and then performs distillation on average representations of these clustered layers.

\subsection{Similarity-based clustering}
Given a small calibration dataset $\mathcal{C}$, we estimate the similarity across teacher layers using the Centered Kernel Alignment (CKA)~\cite{kornblith2019similarity}, a metric commonly used when analyzing neural networks representations. To compute the similarity between two layers $i$ and $j$ we collect the activations of both layers in $\mathbf{X} \in \mathcal{R}^{N \times d_i}$ and $\mathbf{Y} \in \mathcal{R}^{N \times d_j}$, respectively, after processing all N samples from the calibration dataset $\mathcal{C}$. Here, $d_i$ and $d_j$ represent the embedding dimensions. Letting $\mathbf{K} = \mathbf{X}\mathbf{X}^T$ and $\mathbf{L} = \mathbf{Y}\mathbf{Y}^T$, CKA is computed as follows:

\begin{equation}
    CKA(\mathbf{K}, \mathbf{L}) = \frac{HSIC(\mathbf{K}, \mathbf{L})}{\sqrt{HSIC(\mathbf{K},\mathbf{K})HSIC(\mathbf{L},\mathbf{L})}}    
    \label{cka_measure}
\end{equation}

\noindent where HSIC is the Hilbert-Schmidt Independence Criterion \cite{gretton2007kernel}, which is formulated as $HSIC(\mathbf{K}, \mathbf{L}) = \frac{1}{(N-1)^2} tr(\mathbf{K}\mathbf{H}\mathbf{L}\mathbf{H})$,
with $\mathbf{H}= \mathbf{I}_N - \frac{1}{N}\mathbf{1}\mathbf{1}^T$ representing the centering matrix.

By computing the CKA measure for all layers, we construct a similarity matrix as depicted in Fig.~\ref{fig2:summary_method}b. Based on the layer-to-layer similarity, we cluster the teacher layers outputs into a pre-defined number of clusters using agglomerative clustering, a hierarchical clustering method that progressively merges clusters using a bottom-up approach. 

\subsection{Similarity-aware knowledge distillation}
Let $M$ indicate the total number of clusters, then, the distillation loss defined in Eq. \ref{distillation_loss} becomes:

\begin{equation}
    \mathcal{L}^{dis} = \sum_{c=1}^{M} \mathcal{L}(\Bar{\textbf{X}}_c^{\text{tea}}, \Bar{\textbf{X}}_c^{\text{stu}} \cdot \textbf{W}_c)
    \label{our_distillation_loss}
\end{equation}

\noindent where $\Bar{\textbf{X}}_c^{\text{tea}}$ and $\Bar{\textbf{X}}_c^{\text{stu}}$ represent the average representations of cluster $c$ for teacher and student networks, respectively, and $\textbf{W}_c$ is a fully connected layer concurrently trained with the student parameters. We follow DPHuBERT \cite{peng23c_interspeech} in defining the loss function $\mathcal{L}$ as the sum of $\ell_1$ and cosine distances.

Notably, here the layer-wise weights $\gamma_i$ used in Eq.~\ref{distillation_loss} are defined by the clustering step, since the layer-specific influence on the loss is indirectly proportional to the number of layers in the given cluster. In other words, redundant information is automatically weighted down, while unique information is given more importance. 

\section{Experimental setup}
We run our experiments on both WavLM Base+ \cite{chen2022wavlm} and HuBERT Base \cite{hsu2021hubert}, evidencing the adaptability of SKILL to various SSL models. Our experiments are implemented using PyTorch \cite{paszke2019pytorch} and TorchAudio \cite{hwang2022torchaudio}. For comparison, we use the same pre-trained models used in \cite{peng23c_interspeech}, which are downloaded from either fairseq \cite{ott2019fairseq} or Hugging Face \cite{wolf2019huggingface}. The distillation is performed using LibriSpeech 960h ~\cite{panayotov2015librispeech}, while the CKA similarity is computed using a small subset of train-clean 100h. We follow the training recipe of DPHuBERT \cite{peng23c_interspeech}, where each mini-batch contains 640 seconds of audio. The total number of steps in stage one, i.e. joint distillation and structured pruning, is 50k, with 15k warm-up steps. The maximum learning rates are 2e-4 and 2e-2 for main parameters and auxiliary parameters, respectively. The target sparsity is linearly increased in the first 5k steps and is set to 75\% unless differently stated. During the second stage, the total number of steps is 25k, with 5k warm-up steps. The maximum learning rate is 1e-4. 

The evaluation is based on SUPERB \cite{yang2021superb}, which consists of 10 tasks: keyword spotting (KS), intent classification (IC), phoneme recognition (PR), automatic speech recognition (ASR), emotion recognition (ER), query by example spoken term detection (QbE), slot filling (SF), speaker identification (SID), automatic speaker verification (ASV), and speaker diarization (SD). For each task, the SSL model is frozen and a task-specific downstream model is trained using as input a weighted average of the hidden states.

\begin{table}[t!]
    \centering
    \renewcommand{\arraystretch}{0.9}
    \caption{DPWavLM \cite{peng23c_interspeech} performance with varying distillation weights (Eq.~\ref{distillation_loss}). $\gamma$ is the weight used for layers $\{0,4,8,12\}$, weights of other layers are uniformly distributed so that $\sum_{i \in S}\gamma_i=1$.}
    \label{tab:dpwavlm_uniform}
    \small
    \begin{tabular}{ccccccc}
        \toprule
        $\gamma$  & PR & ASR & ER & QbE & SID & SD \\
        &  PER↓ & WER↓ & Acc↑ & MTWV↑ & Acc↑ & DER↓ \\
        \midrule
        1/5 & 7.95 & 10.13 & 65.20 & $\mathbf{8.47}$ & $\mathbf{81.07}$ & 5.42 \\
        1/6  & 7.84 & 10.29 & 65.43 & 8.28 & 80.09 & 5.47 \\
        1/7  & $\mathbf{7.56}$ & $\mathbf{10.09}$ & $\mathbf{65.68}$ & 8.40 & 78.46 & $\mathbf{5.35}$ \\
        1/10  & $\mathbf{7.56}$ & 10.24& 65.53 & 8.07 & 77.25 & 5.63 \\
        1/13  & $\mathbf{7.56}$ & 10.28& 64.80 & 7.47 & 77.57 & 5.72 \\
        \bottomrule
    \end{tabular}
    \vspace{-0.5cm}
\end{table}

\begin{figure}[h]
    \begin{subfigure}{0.2\textwidth}
    \centering
        \includegraphics[width=\textwidth]{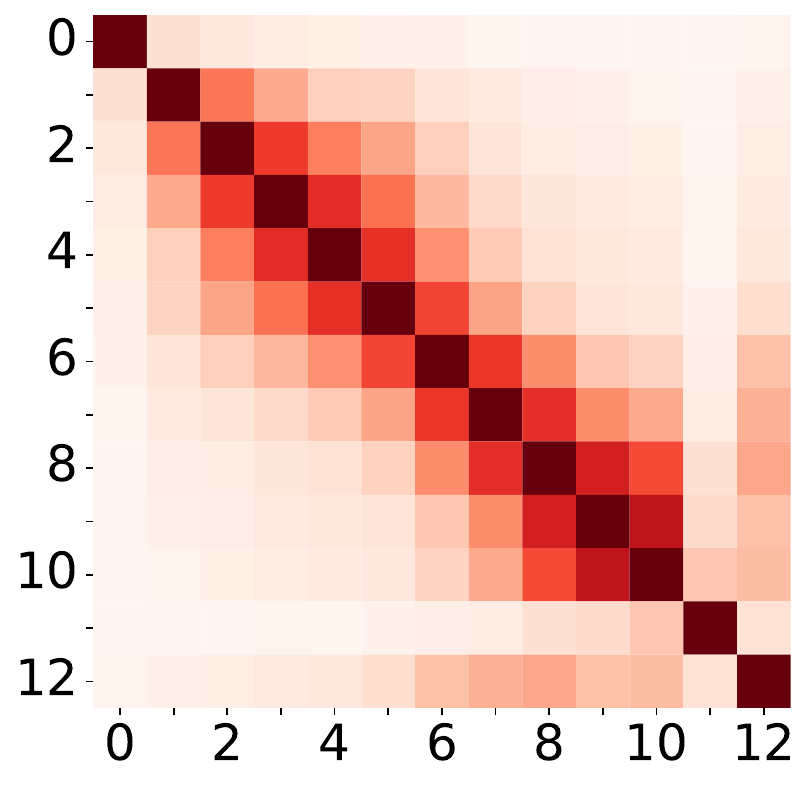}
        \caption{WavLM Base+}
    \end{subfigure}%
     \hfill
    \begin{subfigure}{0.2\textwidth}
    \centering
        \includegraphics[width=\textwidth]{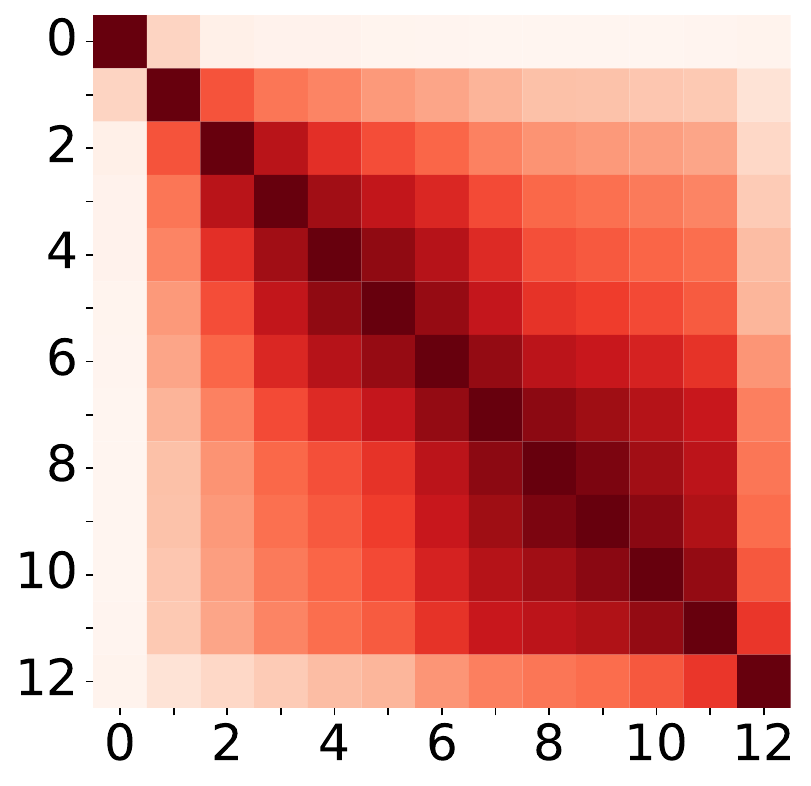}
        \caption{HuBERT Base}
    \end{subfigure}
    \caption{Layer-wise CKA similarity for WavLM Base+ \cite{chen2022wavlm} (a) and HuBERT Base \cite{hsu2021hubert} (b). Darker colors indicate higher similarity.}
    \label{fig2:CKA_similarity}
\vspace{-0.5cm}
\end{figure}

\section{Results}
\label{sec:Results}

\subsection{Limitations of DPHuBERT}
First, we present an analysis that highlights the limitations of the manual layer-selection strategy proposed in \cite{peng23c_interspeech}. Fig.~\ref{fig:DPWavLM_distance} evidences the bias towards the distilled layers in terms of $\ell_1$ and cosine distance, however, it does not relate this pattern to downstream tasks. To do so, we analyze how DPWavLM's performance varies when we gradually increase the relative importance on the distillation loss of layers other than \{0,4,8,12\}. The results are reported in Table \ref{tab:dpwavlm_uniform}.

While the performance on QbE and SID significantly degrades as we move closer to a uniform weighting across all layers, other tasks seem to benefit from it. This highlights the inherent bias of DPHuBERT towards the layers selected to be distilled. Moreover, it should be noticed that no task benefits from distilling all the layers with equal weights (last row in Table \ref{tab:dpwavlm_uniform}). This is likely due to two main aspects: (1) the capacity gap between teacher and student is too large for the student to mimic well all hidden representations. (2) The hidden representations might contain redundant information which should not be distilled. Thus, simply distilling all the layers is not a viable solution to overcome the limitations of DPHuBERT.

\begin{table*}[t]
    \centering
    \renewcommand{\arraystretch}{0.9}
    \setlength{\tabcolsep}{4pt}
    \caption{Comparison of SKILL versus previous distillation methods on SUPERB \cite{yang2021superb}. Our SKHuBERT and SKWavLM are compressed from publicly available HuBERT Base \cite{hsu2021hubert} and WavLM Base+ \cite{chen2022wavlm}  checkpoints, respectively.}
    \label{tab:comparison}
    \small
    \begin{tabular}{lccccccccccc}
        \toprule
        Method & \#Params & KS & IC & PR & ASR & ER & QbE & SF & SID & ASV & SD \\
        & (Millions) & Acc↑ & Acc↑ & PER↓ & WER↓ & Acc↑ & MTWV↑ & F1↑ / CER↓ & Acc↑ & EER↓ & DER↓ \\
        \midrule
        HuBERT Base \cite{hsu2021hubert} & 94.68 & 96.30 & 98.34 & 5.41 & 6.42 & 64.92 & 0.0736 & 88.53 / 25.20 & 81.42 & 5.11 & 5.88 \\
        WavLM Base+ \cite{chen2022wavlm} & 94.70 & 97.37 & 99.00 & 3.92 & 5.59 & 68.65 & 0.0988 & 90.58 / 21.20 & 89.42 & 4.07 & 3.50 \\
        \midrule
        DistilHuBERT \cite{chang2022distilhubert} & 23.49 & 95.98 & 94.99 & 16.27 & 13.37 & 63.02 & 0.0511 & 82.57 / 35.59 & 73.54 & 8.55 & $\mathbf{6.19}$ \\
        FitHuBERT \cite{lee2022fithubert} & 22.49 & 96.27 & 91.25 & 13.32 & 12.09 & 59.82 & 0.0489 & 84.06 / 32.46 & 55.71 & 8.00 & 6.84 \\
        12-Layer Half \cite{ashihara2022deep} & 26.87 & $\mathbf{97.24}$ & 96.97 & 10.67 & 10.96 & 63.24 & 0.0604 & 86.11 / 30.93 & 69.52 & 6.13 & 6.81 \\
        3-Layer One \cite{ashihara2022deep} & 30.58 & 96.69 & 94.15 & 13.34 & 12.23 & $\mathbf{63.95}$ & 0.0489 & 82.89 / 34.65 & 75.71 & 6.48 & 6.56 \\
        ARMHuBERT-S \cite{jang2023interspeech} & 22.39 & 96.82 & 97.02 & 8.63 & 10.82 & 62.96 & $\mathbf{0.0720}$ & 86.34 / 29.02 & 63.76 & $\mathbf{5.58}$ & 7.01 \\
        DPHuBERT \cite{peng23c_interspeech} & 23.58 & 96.33 & 97.33 & 8.85 & $\mathbf{10.51}$ & 63.34 & 0.0692 & 86.07 / 29.15 & $\mathbf{76.46}$ & 6.11 & 6.45 \\
        \textbf{SKHuBERT (ours)} & 23.59 & 96.23 & $\mathbf{97.53}$ & $\mathbf{8.29}$ & 10.78 & 63.83 & 0.0646 & $\mathbf{86.68}$ / $\mathbf{27.45}$ & 74.43 & 6.14 &  6.46 \\
        \midrule
        ARMWavLM-S \cite{jang2023interspeech} & 22.39 & $\mathbf{97.01}$ & 97.76 & 7.42 & $\mathbf{10.03}$ & 64.54 & 0.0741 & 87.41 / 26.97 & 71.29 & 5.99 & 7.11 \\
        DPWavLM \cite{peng23c_interspeech} & 23.58 & 96.23 & $\mathbf{98.66}$ & 7.99 & 10.06 & 65.72 & $\mathbf{0.0865}$ & 88.2 / 26.31 & $\mathbf{80.82}$ & $\mathbf{5.42}$ & 5.58 \\
        \textbf{SKWavLM (ours)} & 23.51 & 96.75 & 98.37 & $\mathbf{6.90}$ & $\mathbf{10.03}$ & $\mathbf{66.22}$ & 0.0727 & $\mathbf{88.62}$ / $\mathbf{25.36}$ & 74.67 & 5.67 & $\mathbf{5.53}$ \\
        \bottomrule
    \end{tabular}
      \vspace{-0.15cm}
\end{table*}

\subsection{SUPERB performance}
As a solution to the aforementioned limitations, we propose SKILL. Table \ref{tab:comparison} compares SKILL with other compression methods. We refer to our compressed WavLM Base+ \cite{chen2022wavlm} and HuBERT Base \cite{hsu2021hubert} models as SKWavLM and SKHuBERT, respectively. We experiment with different clustering settings and present the best performing one, that is, six and eight clusters for WavLM Base+ and HuBERT Base, respectively. We report the performance of DPWavLM and DPHuBERT based on our local implementation, while for the other methods we use publicly available results. SKWavLM outperforms DPWavLM on 6 out of 10 tasks, demonstrating that SKILL leads to better general speech representations, while also simplifying the layer-selection strategy. Notably, SKWavLM exhibits a substantial 27\% reduction in the performance gap with the original WavLM Base+ on PR compared to DPWavLM. Finally, SKWavLM sets new state-of-the-art results on 5 out of 10 tasks in the 30M parameters model class. 
On the other hand, SKHuBERT outperforms DPHuBERT on 4 tasks, with significant improvements on PR and SF. However, we notice that the benefits of SKILL are less evident in this case. We hypothesize that this is due to the higher similarity across layers in HuBERT Base as opposed to WavLM Base+. As depicted in Fig.~\ref{fig2:CKA_similarity}, HuBERT Base exhibits increased redundancy across layers, suggesting that the omission of some layers, as done in DPHuBERT, may not incur as detrimental consequences as observed in WavLM Base+. 

\begin{figure}[h!]
  \centering
  \includegraphics[width=0.47\textwidth]{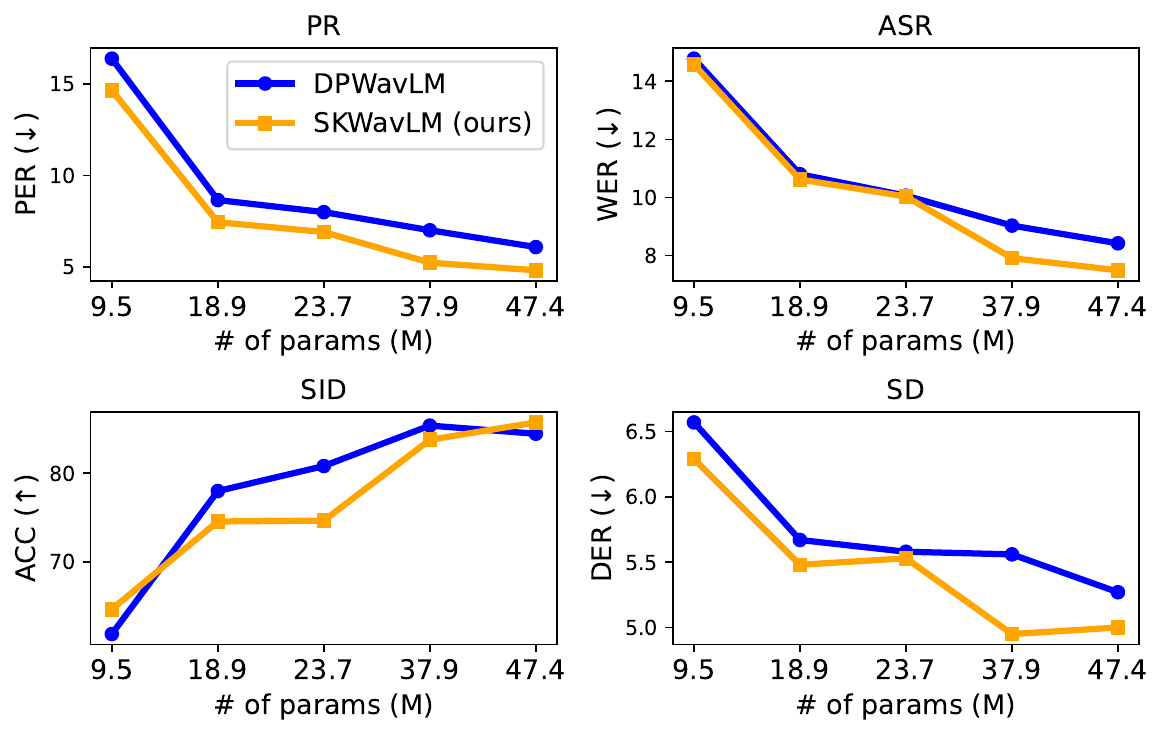}
  \caption{Comparison of SKWavLM with DPWavLM \cite{peng23c_interspeech} on PR, ASR, SID, and SD at different target sparsities.}
  \label{fig:sparsity}
  \vspace{-0.65cm}
\end{figure}

\begin{table*}[ht!]
    \centering
    \renewcommand{\arraystretch}{0.9}
    \setlength{\tabcolsep}{4pt}
    \caption{Ablation study for SKILL. All experiments compress WavLM Base+ \cite{chen2022wavlm} with a target sparsity of 75\%.}
    \label{tab:ablation}
    \small
    \begin{tabular}{lccccccccccc}
        \toprule
        Method & KS & IC & PR & ASR & ER & QbE & SF & SID & ASV & SD \\
        & Acc↑ & Acc↑ & PER↓ & WER↓ & Acc↑ & MTWV↑ & F1↑ / CER↓ & Acc↑ & EER↓ & DER↓ \\
        \midrule
        SKWavLM & $\mathbf{96.75}$ & 98.37 & $\mathbf{6.9}$ & $\mathbf{10.03}$ & $\mathbf{66.22}$ & 0.0727 & $\mathbf{88.62}$ / $\mathbf{25.36}$ & $\mathbf{74.67}$ & 5.67 & 5.53 \\
        w/o averaging  & 96.49 & $\mathbf{98.55}$ & 7.5 & 10.46 & 64.76 & $\mathbf{0.0840}$ & 86.98 / 27.28 & 72.23 & $\mathbf{5.06}$ & $\mathbf{5.49}$ \\
        w/o averaging the student & 96.69 & 98.39 & 7.44 & 10.5 & 64.51 & 0.0745 & 86.29 / 27.94 & 73.47 & 5.85 & 5.50 \\
        \bottomrule
    \end{tabular}
    \vspace{-0.35cm}
\end{table*}

\subsection{Sparsity analysis}
In this section, we assess the robustness of SKILL to various sparsities by training SKWavLM using different targets, as defined in Eq.~\ref{eq:pruning}, and compare it with DPWavLM \cite{peng23c_interspeech}. The results are presented in Fig.4. 
Irrespective of the target network size, SKWavLM outperforms DPWavLM in PR and SD. For ASR, SKWavLM better exploits the additional parameters when reducing sparsity, possibly due to the involvement of all teacher layers during distillation. Lastly, for SID, DPWavLM outperforms SKWavLM when targeting middle-range sparsities, but SKWavLM performs better when targeting tiny (9M parameters) or larger (47.4M parameters) networks, suggesting better adaptability to practical deployments.

\subsection{Ablation study}
The results of the ablation studies are reported in Table 3.

\noindent \textbf{W/o averaging}: SKILL proposes to distill the average representations of the clusters. In this experiment, we perform clustering using the CKA similarity, but the distillation is performed using one layer for each cluster instead of the average. We select the layer with the highest index. While IC, QbE, ASV, and SD benefit from this setting, other tasks show a significant degradation suggesting that averaging helps maintain better generalization capabilities.  

\noindent \textbf{W/o averaging the student}: In SKILL, we align the cluster average representations of teacher and student networks. Here we test whether averaging the student is beneficial by aligning the average cluster representation of the teacher with one layer of the student. For the student, we select the layer with the highest index. The experiments suggest that indeed aligning the average representations is beneficial in most tasks.

\section{Conclusion}
This paper introduces SKILL, a novel group-based distillation method for efficient speech self-supervised model. SKILL extends the joint structured pruning and distillation method proposed in DPHuBERT \cite{peng23c_interspeech}.
Our technique first identifies layers that encode similar information, employing a hierarchical clustering technique based on the layer similarities within the teacher network. The distillation process then exploits the average representations of these identified layers. 
Remarkably, SKILL outperforms DPHuBERT and achieves state-of-the-art results in the 30M parameters model class across several SUPERB tasks, while also demonstrating increased robustness to various sparsity ratios.
We have made our models and distillation recipes publicly available, aiming to contribute to the ongoing efforts in enhancing self-supervised learning through compression techniques. 

\bibliographystyle{IEEEbib}
\bibliography{strings}

\end{document}